\documentclass[aps,prl,reprint,groupedaddress]{revtex4-1}
\usepackage{amsmath}
\usepackage{amssymb}
\usepackage{graphicx}

\begin{document}

%



\title{A Flexible Phase Retrieval Framework for Flux-limited Coherent X-Ray Imaging}

\author{Liang Shi}
\affiliation{Department of Electrical Engineering, Stanford University, Stanford, California 94305, USA}
\author{Gordon Wetzstein}
\affiliation{Department of Electrical Engineering, Stanford University, Stanford, California 94305, USA}
\author{Thomas J. Lane*}
\affiliation{SLAC National Accelerator Laboratory, Menlo Park, California 94025, USA}
\email{tjlane@slac.stanford.edu}
%
%
%

\date{\today}



\begin{abstract}
Coherent X-ray diffraction imaging~(CXDI) experiments are intrinsically limited by shot noise, a lack of knowledge about the sample's support, and missing measurements due to the experimental geometry. We propose a flexible, iterative phase retrieval framework that allows for accurate modeling of Gaussian or Poissonian noise statistics, modified support updates, regularization of  reconstructed signals, and handling of missing data in the observations. The proposed method is efficiently solved using alternating direction method of multipliers~(ADMM) and is demonstrated to consistently outperform state-of-the-art algorithms for low-photon phase retrieval from CXDI experiments, both for simulated diffraction patterns and for experimental measurements.
%
%
%
\end{abstract}


\maketitle

Propelled by the development of ultra-bright $4^{\mathrm{th}}$ generation light sources like the linac coherent light source~(LCLS), coherent X-ray diffraction imaging~(CXDI) has the promise of revealing the atomic or near-atomic structure of aperiodic structures like single, non-crystalline proteins \cite{miao:99,Marchesini:03,thibault:06,seibert:11}.

In a common implementation of such experiments, identically-structured particles are exposed to coherent X-ray pulses at unknown orientations and a detector records the intensity of forward-scattered diffraction patterns. Real-space images can then be retrieved by means of iterative phase retrieval algorithms~\cite{Gerchberg:72,Fienup:82,Bauschke:03,Luke:05,marchesini:07}, for example the hybrid input-output (HIO) and relaxed averaged alternating reflections (RAAR). 

To employ any of these algorithms, it is necessary to  ``oversample'' the diffraction patterns by a factor of at least 2x (in the Shannon sense)~\cite{sayre:52}. Due to this oversampling, reconstructed molecules only occupy part of the image, known as the \emph{support}. Unfortunately, the specific support of any sample is usually unknown and has to be estimated. For example, the Shrinkwrap algorithm~\cite{Marchesini:03}, one of the most common approaches, thresholds the real-space image at each iteration to provide a continuously updated support estimate. An accurate support is crucial for reliable reconstructions.

A second unavoidable challenge in phase retrieval is Poisson-distributed shot noise due to the fact that source brightness or radiation damage intrinsically limits the number of photons a given sample can diffract \cite{Ikeda:12}. Reconstruction resolution is determined by the total number of photons diffracted \cite{Marchesini:03}.
Most iterative phase retrieval algorithms implicitly employ a Gaussian noise model (either as additive readout noise or as a model of the diffraction intensity), which is not optimal for low-light imaging. This model mismatch is small when many photons are captured by each camera pixel ($\gg 10$), but it becomes significant when photons are scarce. Modeling Poisson noise directly with maximum-likelihood methods has demonstrated reconstruction improvement in ptychography \cite{thibault:12} and aberration estimation in incoherent imaging \cite{paxman:92}, suggesting it may also offer advantages in CXDI reconstructions.

Finally, in most CXDI experiments, the direct beam necessitates a beam dump or stop to prevent damage to experimental equipment, resulting in a missing region in the center of the detector that corresponds to low frequency information. This missing information can severely limit the quality of a reconstruction or make it impossible~\cite{Nishino03,Miao05,thibault:06,Huang:10}. Since missing frequencies are unconstrained by available experimental data, they often remain near the random initial values set by the phase retrieval method employed, resulting in significant artifacts in the reconstructed image.  A common empirical solution is to fix the missing intensities to physically motivated values (see \textit{e.g.}~\cite{thibault:06}). Recent studies \cite{Salha:14, He:15}, however, demonstrated that prior information about the real space image, such as total variation regularization (TV)~\cite{rudin:92}, could sufficiently constrain missing intensities, resulting in reproducible, high-quality reconstructions. However, the regularization was carried out in an \emph{ad hoc} manner separate from the phase retrieval algorithm employed (HIO) and therefore was not easily interfaced with other methods.

In this letter, we propose a framework for combining different prior information in an efficient and flexible way, leveraging the alternating direction method of multipliers (ADMM)~\cite{boyd:11}. Similar approaches have been shown successful in phase retrieval for sparse signals \cite{netrapalli:15,weller:15}. Within this framework, we implement a Poisson error model, TV regularization, support update model, and handle missing measurements in a unified manner. Our method is compatible with classic update rules like HIO and RAAR. Here, we focus on the most common case of real and positive images, in both real and diffraction space; extensions to \textit{e.g.}~complex images is straightforward.

%

To be precise, we introduce a notation for CXDI measurements and quickly review these classic phase retrieval algorithms before describing our method mathematically. Let $\mathbf{x}$ be a finite scalar field representing the electron density of an object, generally in 3 dimensions. In this paper, we limit our interest to 2D manifolds within a 3D object and 2D objects. Extension to 3D objects is straight forward. We employ a vectorized representation $\mathbf{x} \in \mathbb{R}^N$, where $x_i$ denotes a voxel value at index $i$, and $i$ runs over all voxels in all dimensions. To deal with missing data in the diffraction image (due to \textit{e.g.}~the beam stop), let $\mathbf{w} \in \mathbb{R}^{N}$ be a binary vector with value 0 for pixels in missing regions and 1 otherwise. In CXDI, the camera measures the diffraction image
\begin{equation*}
	\mathbf{b} = \vert \mathcal{F}\mathbf{x} \vert^2,
\end{equation*}
where $\mathbf{b} \in \mathbb{R}^N$ is the measured Fourier intensity and $\mathcal{F}$ is the 3D discrete Fourier transform operator.  Let $\mathbf{S}^k$ be the \emph{support}, that is $x_i = 0$ if  $i \notin \mathbf{S}^k$, at iteration $k$. Define projection operator $\mathbf{P_m}$,
\begin{equation*}
	\mathbf{P_m}(\mathbf{x}) = \mathcal{F}^{-1}(\mathbf{v}),   \ \ v_{i} = 
	\left\{
	\begin{aligned}
		& \sqrt{b_{i}} \frac{(\mathcal{F} \mathbf{x})_{i}}{|(\mathcal{F} \mathbf{x})_{i}|} \quad W_i = 1, \ |\mathcal{F}\mathbf{x}| \neq 0 \\
		& (\mathcal{F}\mathbf{x})_{i} \qquad \quad \ \ \ \text{otherwise}\\
	\end{aligned}	
	\right.
\end{equation*}
which enforces the amplitude of the diffraction image from the estimated object to equalize the experimental measurement. The Fourier components in missing regions remain unchanged \cite{fienup:93}. 

The three classic algorithms we consider consist of iterative updates, where  in-support pixels are updated according to
\begin{equation*}
	x_{i}^{k+1} = (\mathbf{P_m}\mathbf{x}^{k})_{i}, \quad i \in \mathbf{S}^k,
\end{equation*}
and out-of-support pixels ($i \notin \mathbf{S}^k$) are updated by one of (where the algorithm is indicated)
\begin{equation*}
	x_{i}^{k+1} = 
	\left\{
	\begin{aligned}
		& 0 \qquad \qquad \qquad \qquad \quad \quad \: \ \: Error \ Reduction\\
		& x_{i}^k - \beta (\mathbf{P_m}\mathbf{x}^{k})_{i} \ \qquad \qquad \ \ HIO\\
        & \beta x_{i}^k + (1 - 2\beta) (\mathbf{P_m}\mathbf{x}^{k})_{i} \quad RAAR\\
	\end{aligned}
	\right. 
\end{equation*}
with $\beta$ a feedback parameter (a typical choice of $\beta$ is around $0.9$ \cite{Luke:05}). The update rule for pixels inside the support can be interpreted as a gradient descent update which minimizes 
%
$
	\varepsilon_m = \| \textrm{diag}(\mathbf{w})(\mathcal{F}\mathbf{x} - \sqrt{\mathbf{b}}) \|_2^2,
\label{eq:fourier_error}
$
%
followed by projection onto the feasible set in real space, i.e. support $\mathbf{S}^k$ \cite{marchesini:07}.

The model above assumes the diffracted wavefront can be measured perfectly. A real experiment is better modeled by a Poisson process that takes into account photon shot statistics
\begin{equation*}
\mathbf{b} \sim \mathcal{P} \left( |\mathcal{F} \mathbf{x}|^2 \right).
\end{equation*}
specifically, the probability of observing $b_{i}$ photons at pixel $i$ is
\begin{equation*}
p \left( b_{i} | \mathbf{x} \right) =  
\left\{
\begin{aligned}
\frac{ \left(|\mathcal{F} \mathbf{x}|^2 \right)_{i}^{b_{i}} e^{-\left(|\mathcal{F} \mathbf{x}|^2 \right)_{i}} }{b_{i}!}, \quad & W_{i} = 1\\
const \qquad \qquad, \quad & W_{i} = 0
\end{aligned}
\right.
\end{equation*}
Let $\mathbf{1}$ be a column vector with every element 1, the log-likelihood of the joint probability can then be expressed in following compact form
\begin{equation*}
L_{\mathbf{W}} \left( \mathbf{x} \right) = \left(\mathbf{W} \, \textrm{log} |\mathcal{F} \mathbf{x}|^2 \right)^T \mathbf{b} - \left( \mathbf{W} |\mathcal{F} \mathbf{x}|^2 \right)^T \mathbf{1} - \sum_{i=1}^{N} W_i \textrm{log} \left( b_{i} ! \right)
\end{equation*}
where $\mathbf{W} = \textrm{diag}(\mathbf{w})$, with gradient
\vspace{-1 mm}
\begin{equation*}
\nabla L_{\mathbf{W}} \left( \mathbf{x} \right) = 2\mathcal{F}^{-1}\bigg(\mathcal{F} \mathbf{x} - \mathbf{W} \, \Big(\textrm{diag}({|\mathcal{F} \mathbf{x}|^2})^{-1}\textrm{diag}(\mathcal{F} \mathbf{x})\mathbf{b}\Big)\bigg)
\end{equation*}
allowing the log-likelihood to be efficiently maximized by gradient ascent.
\begin{figure}[t]
\centering\includegraphics[width=1.0\linewidth]{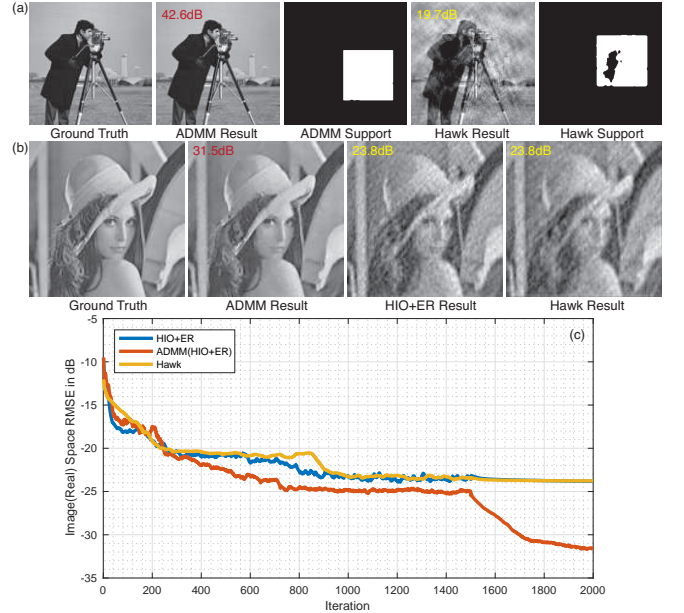}
\caption{(a) Reconstructed ``Cameraman'' and associated supports from 2.5x oversampled noise-free simulations, for both our method (ADMM) and the Hawk package using $\sigma_{init} = 3$, $\sigma_{end} = 1.5$. PSNRs labeled on top-left corner of each image. (b) Reconstructed ``Lena'' from 2x oversampled simulations with shot-noise (average 150 photons/px), for our method (ADMM), HIO and ER implemented by the authors, and the Hawk package using $\sigma_{init} = 1$, $\sigma_{end} = 0.5$. PSNRs labeled on top-left corner of each image. (c) Real-space root-mean-square error (RMSE) of ``Lena'' between the ground truth and reconstruction over 2000 iterations.}
\label{fig:STI_fg1}
\hrulefill
\vspace{-3 mm}
\end{figure}

We add two pieces of prior knowledge to this model of the measurement: that we expect the real space image to be smooth and positive. This is not an exhaustive list of prior information that could be employed, but will be used to demonstrate the effectiveness of our approach. We add prior terms directly to the objective function, allowing for continuous enforcement of this information. No intensity constraints are applied to missing-measurement regions, though these regions are implicitly constrained by the real-space priors. We can then write an objective function for the phase retrieval problem with these priors,
%

\vspace{-1 mm}
\begin{equation}
\underset{\mathbf{x}}{\textrm{minimize}} \quad - L_{\mathbf{W}} \left( \mathbf{x} \right) + \lambda \left\| \mathbf{Dx} \right\|_{2,1} + \mathcal{I}_{\mathbb{R}_+} \left( \mathbf{x} \right),
\label{eq:obj_raw}
\end{equation}
where $\mathbf{D} = [\mathbf{D}_x;\mathbf{D}_y] \in \mathbb{R}^{2N \times N}$ is the discrete gradient operator consisting the horizontal and vertical partial derivative operators(vertically stacked), $\|\mathbf{Dx}\|_{2,1} = \sum\nolimits_{g\in\{x,y\}} \|\mathbf{D}_g\mathbf{x}\|_2$ is the $\ell_1/\ell_2$ norm of $\mathbf{Dx}$, $\lambda$ is the weight for the isotropic TV term and $\mathcal{I}_{\mathbb{R}_+}(\mathbf{x})$ is the indicator function 
%
enforcing positivity ($\mathcal{I}_{\mathbb{R}_+} = 0$ if $x_i \in \mathbb{R}_+$ and $\infty$ otherwise). During iteration, $\mathcal{I}_{\mathbb{R}_+}$ is not evaluated in the objective, only in the gradient.

This problem can be efficiently solved by the alternating direction of multipler methods (ADMM)~\cite{boyd:11}. Using ADMM, Eq.~(\ref{eq:obj_raw}) is reformulated as
%

\begin{align}
\label{eq:ADMMobjective}
\underset{\mathbf{x}}{\textrm{minimize}} & \quad \underbrace{- L_{\mathbf{W}}}_{f \left( \mathbf{x} \right)} + \underbrace{\lambda \left\| \mathbf{z}_1 \right\|_{2,1}}_{g_1 \left( \mathbf{z}_1 \right)} + \underbrace{\mathcal{I}_{\mathbb{R}_+}(\mathbf{z}_2)}_{g_2 \left( \mathbf{z}_2 \right)} \\
\textrm{subject to} & \quad \underbrace{\left[ 
\begin{array}{c}
\mathbf{D}  \\
\mathbf{I}
\end{array}
\right]}_{\mathbf{K}} \mathbf{x} - \underbrace{\left[ 
\begin{array}{c}
\mathbf{z}_1  \\
\mathbf{z}_2
\end{array}
\right]}_{\mathbf{z}} = 0,
\label{eq:ADMM_constraint}
\end{align}
where $\mathbf{z}_1 \in \mathbb{R}^{2N}$ and $\mathbf{z}_2 \in \mathbb{R}^{N}$ are slack variables. ADMM splits the objective into a weighted sum of three independent functions $f(\mathbf{x})$, $g_1(\mathbf{z}_1)$ and $g_2(\mathbf{z}_2)$ that are only linked through the stated constraints.
%
Following the general ADMM strategy, we write an augmented Lagrangian of Eq.~(\ref{eq:ADMMobjective})
\begin{align}
L_{\rho_1,\rho_2} \left( \mathbf{x}, \mathbf{z}, \mathbf{y} \right) = & f \left( \mathbf{x} \right) + g_1 \left( \mathbf{z}_1 \right) + g_2 \left( \mathbf{z}_2\right) + \mathbf{y}^T \left( \mathbf{K} \mathbf{x} - \mathbf{z} \right)\nonumber\\
& + \frac{\rho_1}{2} \left\| \mathbf{D} \mathbf{x} - \mathbf{z}_1 \right\|_2^2 + \frac{\rho_2}{2} \left\| \mathbf{x} - \mathbf{z}_2 \right\|_2^2,
\label{eq:augmented_obj}
\end{align}
where  $\rho_1$ and $\rho_2$ set the penalty associated with ADMM constraints violation. Under the scaled form of the augmented Lagrangian, a single ADMM iteration consists of a sequential updates:
\begin{align}
	\mathbf{x} & \leftarrow \mathbf{prox}_{quad,\rho_1,\rho_2} \left( \mathbf{v_1,v_2} \right) = \underset{\mathbf{x}}{\textrm{arg min}} \ f \left( \mathbf{x} \right)  +\nonumber \\
	& \frac{\rho_1}{2} \left\| \mathbf{D} \mathbf{x} - \mathbf{v}_1 \right\|_2^2 \quad + \frac{\rho_2}{2} \left\| \mathbf{x} - \mathbf{v}_2 \right\|_2^2,\nonumber \\
	 & \quad \mathbf{v}_1 = \mathbf{z}_1 - \mathbf{u}_1, \ \mathbf{v}_2 = \mathbf{z}_2 - \mathbf{u}_2 \nonumber \nonumber \\
	\mathbf{z}_1 & \leftarrow \mathbf{prox}_{\left\| \cdot \right\|_1,\rho_1} \left( \mathbf{v} \right) = \underset{{\mathbf{z}_1}}{\textrm{arg min}} \ g_1 \left( \mathbf{z}_1 \right) + \frac{\rho_1}{2} \left\| \mathbf{v} - \mathbf{z}_1 \right\|_2^2,\nonumber\\ 
    & \quad \ \mathbf{v} = \mathbf{D} \mathbf{x} + \mathbf{u}_1 \nonumber \\
	\mathbf{z}_2 & \leftarrow \mathbf{prox}_{\mathcal{I},\rho_2} \left( \mathbf{v} \right) = \underset{{\mathbf{z}_2}}{\textrm{arg min}} \ g_2 \left( \mathbf{z}_2 \right) + \frac{\rho_2}{2} \left\| \mathbf{v} - \mathbf{z}_2 \right\|_2^2, \nonumber\\
    & \quad \ \mathbf{v} = \mathbf{x} + \mathbf{u}_2 \nonumber\\
	\underbrace{\left[ 
		\begin{array}{c}
		\mathbf{u}_1  \\
		\mathbf{u}_2
		\end{array}
		\right]}_{\mathbf{u}} & \leftarrow \mathbf{u} + \mathbf{K} \mathbf{x} - \mathbf{z} ,
	\label{eq:ADMMupdateruless}
\end{align}
where the scaled dual variable $\mathbf{u} = (1 / \rho) \mathbf{y}$ is used to simplify the notation of Eq.~(\ref{eq:augmented_obj}).

The $\mathbf{x}$-update is a quadratic program which can be iteratively minimized by gradient-based methods. We used the Hessian-free Newton method provided by the minFunc package (http://www.cs.ubc.ca/\textasciitilde schmidtm/Software/minFunc.html). Upon completion, out-of-support pixels can be updated using rules specified by HIO or RAAR.

After $t$ ADMM iterations, we update the object support using a modified Shrinkwrap algorithm. Let $G(k_r)$ be a normalized Gaussian blurring kernel with standard deviation $k_r$, and $\eta$ be a thresholding parameter. Standard Shrinkwrap updates the support by convolving the realspace image with a Gaussian and thresholding, precisely
\begin{enumerate}
	\item $\mathbf{x_g} = |\mathbf{x}|*G(k_r)$
    \item $\mathbf{S} = \left\{ i \ | \ (\mathbf{x_g})_{i} \geq \eta\max(\mathbf{x_g}) \right\}$. 
\end{enumerate}
Inspired by recent study which showed that modest overestimation of the extent of the support results in significantly less error than underestimating it \cite{Huang:10}, we added a third step that fills in any non-support regions completely enclosed by the support. Specifically, we employed a morphological hole-filling on the binary support image \cite{Soille:99}. We also experimented with including the entire convex hull of the Shrinkwrap support, with good but inferior results.
\begin{figure}[t]
\centering\includegraphics[width=1.0\linewidth]{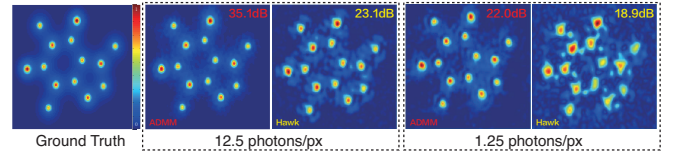}
\caption{Reconstructed caffeine molecule from 2x oversampled simulations with shot-noise at average [12.5, 1.25] photons/px, for our method (ADMM) and the Hawk package using $\sigma_{init} = 1$, $\sigma_{end} = 0.5$. PSNRs labeled on top-right corner of each image.}
\label{fig:PDB_fg1}
\hrulefill
\vspace{-5 mm}
\end{figure}

For the $\mathbf{z}_1$ and $\mathbf{z}_2$ update steps, closed-form solutions can be obtained using proximate operators for the $\ell_1/\ell_2$-norm and indicator function, as discussed in~\cite{parikh:14}.

The ADMM framework allows flexible expression of models that can be customized easily and solved efficiently. For example, switching to a Gaussian error model only requires modification on the cost function of the $\mathbf{x}$-update and its gradient. Because the slack variable~($\mathbf{z}$) update is separated from the unknown variable~($\mathbf{x}$) update, changing or imposing new priors is convenient.

To assess the proposed algorithm, we applied it to both simulated diffraction patterns and experimental measurement imaged at LCLS. The results are compared with the ones produced by the state-of-the-art phase retrieval toolbox Hawk~\cite{maia:10}, which implements the standard shrinkwrap algorithm, HIO and RAAR with a Gaussian noise model (but not a Poisson noise model).


We began with test images (Figures \ref{fig:STI_fg1} and \ref{fig:PDB_fg1}). Initial supports were obtained by thresholding the inverse Fourier transform of intensity measurement (autocorrelation) at 4\% of the maximum intensity. The initial and minimal standard deviation of Shrinkwrap Gaussian blur kernel($\sigma_{init},\sigma_{min}$) were chosen based on trails and the initial value was reduced by 1\% every 20 iterations down to the minimum. The  parameters $\eta = [5\%, \ 10\%, \ 15\%, \ 20\%]$ and $\lambda = [0.1, \ 0.01, \ 0.005, \ 0.001]$ were tested and the resulting highest peak signal-to-noise ratio~(PSNR) reconstruction was chosen. The slack variables were set to $\rho_1 = 50 \lambda / \max (\mathbf{x})$ and $\rho_2 = \rho_1/100$ by empirical trial and error. A Gaussian noise model was used to reconstruct noise-free simulations while the Poisson model was used for simulations incorporating shot-noise. All reconstructions (both Hawk and ADMM) were run for 1500 global iterations using HIO update with $\beta = 0.9$. For 500 additional iterations, the out-of-support update was switched from HIO to ER. The $\mathbf{x}$-update ADMM step ran for 20 internal iterations in each global ADMM iteration. Reconstructions were rotated to appear "upright" in the case that the algorithm produced an upside-down image, which is expected to occur in half of randomly seeded reconstructions due to the inversion (Friedel) symmetry of real diffraction images.

Fig.~\ref{fig:STI_fg1}a demonstrates the importance of support estimation. The low brightness of the camera man's black coat produces holes in the support estimated by Hawk, whereas our modified Shrinkwrap fills these in, leading to a better reconstruction. Fig.~\ref{fig:STI_fg1}b, showing reconstructions of the ``Lena'' from simulations incorporating modest shot-noise, demonstrates how the Poisson model and prior constraints jointly improve the final reconstruction. Fig.~\ref{fig:STI_fg1}c shows how the error between the reconstruction and the ground truth improves during algorithm iterations. Our method converges in a comparable number of iterations to Hawk, but typically finds lower-error solutions.

%
%

For a more experimentally relevant test, we simulated diffraction from a caffeine molecule's electron density~\cite{berman:02}. Fig.~\ref{fig:PDB_fg1} shows the reconstructed results from measurements at different shot-noise levels. As shot noise increases, Hawk increasingly suffers from artifacts, where our method faithfully preserves the shape of molecule even with 10 times fewer photons. 

Finally, we assess the proposed algorithm on experimentally measured mimivirus diffraction patterns obtained at LCLS \cite{seibert:11, maia:12}. The $512 \times 512$~px measurement contains a missing sphere of radius $35$~px and a $20$~px tall horizontal missing slit across the center. The RAAR algorithm was applied to allow comparison with published results \cite{seibert:11}. We employed the area-based Shrinkwrap algorithm and RAAR parameters used in that work.

\begin{figure}[t]
\centering\includegraphics[width=1.0\linewidth]{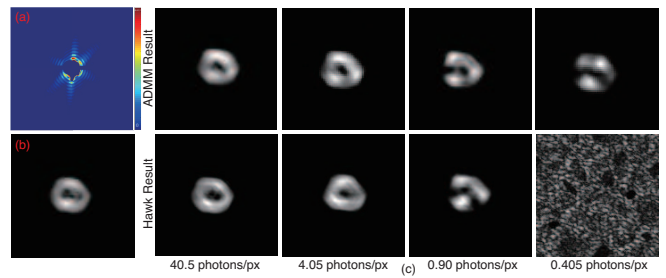}
\caption{(a) Experimentally recorded far-field diffraction pattern of a single mimivirus. (b) Reconstructed mimivirus from the original measurement using Hawk's implementation of RAAR. (c) Reconstructed mimivirus from shot-noise simulations at average [40.5, 4.05, 0.90, 0.405] photons/px, for our method (ADMM) and the Hawk package using $\sigma_{init} = 10$, $\sigma_{end} = 1$.}
\label{fig:mimi_fg1}
\hrulefill
\vspace{-5 mm}
\end{figure}

Fig.~\ref{fig:mimi_fg1} shows the reconstructed mimivirus from simulations at different shot-noise levels, where we modeled the original measurement as noise-free ground-truth and sampled it with Poisson statistics to mimic the diffraction from a smaller object or weaker source. Employing Hawk's RAAR implementation, a drop in reconstruction quality occurs at $0.90$ photons/px, and the virus becomes completely irretrievable at $0.405$ photons/px. ADMM RAAR preserves the approximate shape of virus at $0.405$ photons/px, with one missing and one attenuated edge of the hexagon-like shape of the projected virus.

%
%
%
%
In conclusion, we propose a phase retrieval framework that allows simultaneous optimization of both the primary model (here, Poisson or Gaussian pixel noise) and the prior constraints, with proper handling of missing measurements. It is efficiently solved using ADMM, by splitting the objective into sub-problems and addressing them independently. The decoupled treatment and the flexibility of adding/dropping priors at run time provides a significant productivity advantage. Together with a modified Shrinkwrap support update, the proposed algorithm produces high-quality reconstructions, in particular for CXDI measurements with low photon counts.
\begin{acknowledgments}
This work was funded through the LCLS Directorate of SLAC National Accelerator Laboratory under DoE BES Contract No. DE-AC02-76SF00515 (LS and TL) and an NSF CAREER Award No. IIS 155333 (GW). Thanks to Daniel Ratner and Filipe Maia for comments on a draft. TL would like to acknowledge Sara Salha and Kevin Raines for sharing thoughts on Bayesian phase retrieval (as in eq.~\ref{eq:obj_raw}, manuscript forthcoming).
\end{acknowledgments}


\begin{thebibliography}{28}%
\makeatletter
\providecommand \@ifxundefined [1]{%
 \@ifx{#1\undefined}
}%
\providecommand \@ifnum [1]{%
 \ifnum #1\expandafter \@firstoftwo
 \else \expandafter \@secondoftwo
 \fi
}%
\providecommand \@ifx [1]{%
 \ifx #1\expandafter \@firstoftwo
 \else \expandafter \@secondoftwo
 \fi
}%
\providecommand \natexlab [1]{#1}%
\providecommand \enquote  [1]{``#1''}%
\providecommand \bibnamefont  [1]{#1}%
\providecommand \bibfnamefont [1]{#1}%
\providecommand \citenamefont [1]{#1}%
\providecommand \href@noop [0]{\@secondoftwo}%
\providecommand \href [0]{\begingroup \@sanitize@url \@href}%
\providecommand \@href[1]{\@@startlink{#1}\@@href}%
\providecommand \@@href[1]{\endgroup#1\@@endlink}%
\providecommand \@sanitize@url [0]{\catcode `\\12\catcode `\$12\catcode
  `\&12\catcode `\#12\catcode `\^12\catcode `\_12\catcode `\%12\relax}%
\providecommand \@@startlink[1]{}%
\providecommand \@@endlink[0]{}%
\providecommand \url  [0]{\begingroup\@sanitize@url \@url }%
\providecommand \@url [1]{\endgroup\@href {#1}{\urlprefix }}%
\providecommand \urlprefix  [0]{URL }%
\providecommand \Eprint [0]{\href }%
\providecommand \doibase [0]{http://dx.doi.org/}%
\providecommand \selectlanguage [0]{\@gobble}%
\providecommand \bibinfo  [0]{\@secondoftwo}%
\providecommand \bibfield  [0]{\@secondoftwo}%
\providecommand \translation [1]{[#1]}%
\providecommand \BibitemOpen [0]{}%
\providecommand \bibitemStop [0]{}%
\providecommand \bibitemNoStop [0]{.\EOS\space}%
\providecommand \EOS [0]{\spacefactor3000\relax}%
\providecommand \BibitemShut  [1]{\csname bibitem#1\endcsname}%
\let\auto@bib@innerbib\@empty
\bibitem [{\citenamefont {Miao}\ \emph {et~al.}(1999)\citenamefont {Miao},
  \citenamefont {Charalambous}, \citenamefont {Kirz},\ and\ \citenamefont
  {Sayre}}]{miao:99}%
  \BibitemOpen
  \bibfield  {author} {\bibinfo {author} {\bibfnamefont {J.}~\bibnamefont
  {Miao}}, \bibinfo {author} {\bibfnamefont {P.}~\bibnamefont {Charalambous}},
  \bibinfo {author} {\bibfnamefont {J.}~\bibnamefont {Kirz}}, \ and\ \bibinfo
  {author} {\bibfnamefont {D.}~\bibnamefont {Sayre}},\ }\href@noop {}
  {\bibfield  {journal} {\bibinfo  {journal} {Nature}\ }\textbf {\bibinfo
  {volume} {400}},\ \bibinfo {pages} {342} (\bibinfo {year}
  {1999})}\BibitemShut {NoStop}%
\bibitem [{\citenamefont {Marchesini}\ \emph {et~al.}(2003)\citenamefont
  {Marchesini}, \citenamefont {He}, \citenamefont {Chapman}, \citenamefont
  {Hau-Riege}, \citenamefont {Noy}, \citenamefont {Howells}, \citenamefont
  {Weierstall},\ and\ \citenamefont {Spence}}]{Marchesini:03}%
  \BibitemOpen
  \bibfield  {author} {\bibinfo {author} {\bibfnamefont {S.}~\bibnamefont
  {Marchesini}}, \bibinfo {author} {\bibfnamefont {H.}~\bibnamefont {He}},
  \bibinfo {author} {\bibfnamefont {H.~N.}\ \bibnamefont {Chapman}}, \bibinfo
  {author} {\bibfnamefont {S.~P.}\ \bibnamefont {Hau-Riege}}, \bibinfo {author}
  {\bibfnamefont {A.}~\bibnamefont {Noy}}, \bibinfo {author} {\bibfnamefont
  {M.~R.}\ \bibnamefont {Howells}}, \bibinfo {author} {\bibfnamefont
  {U.}~\bibnamefont {Weierstall}}, \ and\ \bibinfo {author} {\bibfnamefont
  {J.~C.~H.}\ \bibnamefont {Spence}},\ }\href {\doibase
  10.1103/PhysRevB.68.140101} {\bibfield  {journal} {\bibinfo  {journal} {Phys.
  Rev. B}\ }\textbf {\bibinfo {volume} {68}},\ \bibinfo {pages} {140101}
  (\bibinfo {year} {2003})}\BibitemShut {NoStop}%
\bibitem [{\citenamefont {Thibault}\ \emph {et~al.}(2006)\citenamefont
  {Thibault}, \citenamefont {Elser}, \citenamefont {Jacobsen}, \citenamefont
  {Shapiro},\ and\ \citenamefont {Sayre}}]{thibault:06}%
  \BibitemOpen
  \bibfield  {author} {\bibinfo {author} {\bibfnamefont {P.}~\bibnamefont
  {Thibault}}, \bibinfo {author} {\bibfnamefont {V.}~\bibnamefont {Elser}},
  \bibinfo {author} {\bibfnamefont {C.}~\bibnamefont {Jacobsen}}, \bibinfo
  {author} {\bibfnamefont {D.}~\bibnamefont {Shapiro}}, \ and\ \bibinfo
  {author} {\bibfnamefont {D.}~\bibnamefont {Sayre}},\ }\href@noop {}
  {\bibfield  {journal} {\bibinfo  {journal} {Acta Crystallographica Section A:
  Foundations of Crystallography}\ }\textbf {\bibinfo {volume} {62}},\ \bibinfo
  {pages} {248} (\bibinfo {year} {2006})}\BibitemShut {NoStop}%
\bibitem [{\citenamefont {Seibert}\ \emph {et~al.}(2011)\citenamefont
  {Seibert}, \citenamefont {Ekeberg}, \citenamefont {Maia}, \citenamefont
  {Svenda}, \citenamefont {Andreasson}, \citenamefont {J{\"o}nsson},
  \citenamefont {Odi{\'c}}, \citenamefont {Iwan}, \citenamefont {Rocker},
  \citenamefont {Westphal}, \citenamefont {Hantke}, \citenamefont {DePonte},
  \citenamefont {Barty}, \citenamefont {Schulz}, \citenamefont {Gumprecht},
  \citenamefont {Coppola}, \citenamefont {Aquila}, \citenamefont {Liang},
  \citenamefont {White}, \citenamefont {Martin}, \citenamefont {Caleman},
  \citenamefont {Stern}, \citenamefont {Abergel}, \citenamefont {Seltzer},
  \citenamefont {Claverie}, \citenamefont {Bostedt}, \citenamefont {Bozek},
  \citenamefont {Boutet}, \citenamefont {Miahnahri}, \citenamefont
  {Messerschmidt}, \citenamefont {Krzywinski}, \citenamefont {Williams},
  \citenamefont {Hodgson}, \citenamefont {Bogan}, \citenamefont {Hampton},
  \citenamefont {Sierra}, \citenamefont {Starodub}, \citenamefont {Andersson},
  \citenamefont {Bajt}, \citenamefont {Barthelmess}, \citenamefont {Spence},
  \citenamefont {Fromme}, \citenamefont {Weierstall}, \citenamefont {Kirian},
  \citenamefont {Hunter}, \citenamefont {Doak}, \citenamefont {Marchesini},
  \citenamefont {Hau-Riege}, \citenamefont {Frank}, \citenamefont {Shoeman},
  \citenamefont {Lomb}, \citenamefont {Epp}, \citenamefont {Hartmann},
  \citenamefont {Rolles}, \citenamefont {Rudenko}, \citenamefont {Schmidt},
  \citenamefont {Foucar}, \citenamefont {Kimmel}, \citenamefont {Holl},
  \citenamefont {Rudek}, \citenamefont {Erk}, \citenamefont {H{\"o}mke},
  \citenamefont {Reich}, \citenamefont {Pietschner}, \citenamefont
  {Weidenspointner}, \citenamefont {Str{\"u}der}, \citenamefont {Hauser},
  \citenamefont {Gorke}, \citenamefont {Ullrich}, \citenamefont {Schlichting},
  \citenamefont {Herrmann}, \citenamefont {Schaller}, \citenamefont {Schopper},
  \citenamefont {Soltau}, \citenamefont {K{\"u}hnel}, \citenamefont
  {Andritschke}, \citenamefont {Schr{\"o}ter}, \citenamefont {Krasniqi},
  \citenamefont {Bott}, \citenamefont {Schorb}, \citenamefont {Rupp},
  \citenamefont {Adolph}, \citenamefont {Gorkhover}, \citenamefont {Hirsemann},
  \citenamefont {Potdevin}, \citenamefont {Graafsma}, \citenamefont {Nilsson},
  \citenamefont {Chapman},\ and\ \citenamefont {Hajdu}}]{seibert:11}%
  \BibitemOpen
  \bibfield  {author} {\bibinfo {author} {\bibfnamefont {M.~M.}\ \bibnamefont
  {Seibert}}, \bibinfo {author} {\bibfnamefont {T.}~\bibnamefont {Ekeberg}},
  \bibinfo {author} {\bibfnamefont {F.~R. N.~C.}\ \bibnamefont {Maia}},
  \bibinfo {author} {\bibfnamefont {M.}~\bibnamefont {Svenda}}, \bibinfo
  {author} {\bibfnamefont {J.}~\bibnamefont {Andreasson}}, \bibinfo {author}
  {\bibfnamefont {O.}~\bibnamefont {J{\"o}nsson}}, \bibinfo {author}
  {\bibfnamefont {D.}~\bibnamefont {Odi{\'c}}}, \bibinfo {author}
  {\bibfnamefont {B.}~\bibnamefont {Iwan}}, \bibinfo {author} {\bibfnamefont
  {A.}~\bibnamefont {Rocker}}, \bibinfo {author} {\bibfnamefont
  {D.}~\bibnamefont {Westphal}}, \bibinfo {author} {\bibfnamefont
  {M.}~\bibnamefont {Hantke}}, \bibinfo {author} {\bibfnamefont {D.~P.}\
  \bibnamefont {DePonte}}, \bibinfo {author} {\bibfnamefont {A.}~\bibnamefont
  {Barty}}, \bibinfo {author} {\bibfnamefont {J.}~\bibnamefont {Schulz}},
  \bibinfo {author} {\bibfnamefont {L.}~\bibnamefont {Gumprecht}}, \bibinfo
  {author} {\bibfnamefont {N.}~\bibnamefont {Coppola}}, \bibinfo {author}
  {\bibfnamefont {A.}~\bibnamefont {Aquila}}, \bibinfo {author} {\bibfnamefont
  {M.}~\bibnamefont {Liang}}, \bibinfo {author} {\bibfnamefont {T.~A.}\
  \bibnamefont {White}}, \bibinfo {author} {\bibfnamefont {A.}~\bibnamefont
  {Martin}}, \bibinfo {author} {\bibfnamefont {C.}~\bibnamefont {Caleman}},
  \bibinfo {author} {\bibfnamefont {S.}~\bibnamefont {Stern}}, \bibinfo
  {author} {\bibfnamefont {C.}~\bibnamefont {Abergel}}, \bibinfo {author}
  {\bibfnamefont {V.}~\bibnamefont {Seltzer}}, \bibinfo {author} {\bibfnamefont
  {J.-M.}\ \bibnamefont {Claverie}}, \bibinfo {author} {\bibfnamefont
  {C.}~\bibnamefont {Bostedt}}, \bibinfo {author} {\bibfnamefont {J.~D.}\
  \bibnamefont {Bozek}}, \bibinfo {author} {\bibfnamefont {S.}~\bibnamefont
  {Boutet}}, \bibinfo {author} {\bibfnamefont {A.~A.}\ \bibnamefont
  {Miahnahri}}, \bibinfo {author} {\bibfnamefont {M.}~\bibnamefont
  {Messerschmidt}}, \bibinfo {author} {\bibfnamefont {J.}~\bibnamefont
  {Krzywinski}}, \bibinfo {author} {\bibfnamefont {G.}~\bibnamefont
  {Williams}}, \bibinfo {author} {\bibfnamefont {K.~O.}\ \bibnamefont
  {Hodgson}}, \bibinfo {author} {\bibfnamefont {M.~J.}\ \bibnamefont {Bogan}},
  \bibinfo {author} {\bibfnamefont {C.~Y.}\ \bibnamefont {Hampton}}, \bibinfo
  {author} {\bibfnamefont {R.~G.}\ \bibnamefont {Sierra}}, \bibinfo {author}
  {\bibfnamefont {D.}~\bibnamefont {Starodub}}, \bibinfo {author}
  {\bibfnamefont {I.}~\bibnamefont {Andersson}}, \bibinfo {author}
  {\bibfnamefont {S.}~\bibnamefont {Bajt}}, \bibinfo {author} {\bibfnamefont
  {M.}~\bibnamefont {Barthelmess}}, \bibinfo {author} {\bibfnamefont
  {J.~C.~H.}\ \bibnamefont {Spence}}, \bibinfo {author} {\bibfnamefont
  {P.}~\bibnamefont {Fromme}}, \bibinfo {author} {\bibfnamefont
  {U.}~\bibnamefont {Weierstall}}, \bibinfo {author} {\bibfnamefont
  {R.}~\bibnamefont {Kirian}}, \bibinfo {author} {\bibfnamefont
  {M.}~\bibnamefont {Hunter}}, \bibinfo {author} {\bibfnamefont {R.~B.}\
  \bibnamefont {Doak}}, \bibinfo {author} {\bibfnamefont {S.}~\bibnamefont
  {Marchesini}}, \bibinfo {author} {\bibfnamefont {S.~P.}\ \bibnamefont
  {Hau-Riege}}, \bibinfo {author} {\bibfnamefont {M.}~\bibnamefont {Frank}},
  \bibinfo {author} {\bibfnamefont {R.~L.}\ \bibnamefont {Shoeman}}, \bibinfo
  {author} {\bibfnamefont {L.}~\bibnamefont {Lomb}}, \bibinfo {author}
  {\bibfnamefont {S.~W.}\ \bibnamefont {Epp}}, \bibinfo {author} {\bibfnamefont
  {R.}~\bibnamefont {Hartmann}}, \bibinfo {author} {\bibfnamefont
  {D.}~\bibnamefont {Rolles}}, \bibinfo {author} {\bibfnamefont
  {A.}~\bibnamefont {Rudenko}}, \bibinfo {author} {\bibfnamefont
  {C.}~\bibnamefont {Schmidt}}, \bibinfo {author} {\bibfnamefont
  {L.}~\bibnamefont {Foucar}}, \bibinfo {author} {\bibfnamefont
  {N.}~\bibnamefont {Kimmel}}, \bibinfo {author} {\bibfnamefont
  {P.}~\bibnamefont {Holl}}, \bibinfo {author} {\bibfnamefont {B.}~\bibnamefont
  {Rudek}}, \bibinfo {author} {\bibfnamefont {B.}~\bibnamefont {Erk}}, \bibinfo
  {author} {\bibfnamefont {A.}~\bibnamefont {H{\"o}mke}}, \bibinfo {author}
  {\bibfnamefont {C.}~\bibnamefont {Reich}}, \bibinfo {author} {\bibfnamefont
  {D.}~\bibnamefont {Pietschner}}, \bibinfo {author} {\bibfnamefont
  {G.}~\bibnamefont {Weidenspointner}}, \bibinfo {author} {\bibfnamefont
  {L.}~\bibnamefont {Str{\"u}der}}, \bibinfo {author} {\bibfnamefont
  {G.}~\bibnamefont {Hauser}}, \bibinfo {author} {\bibfnamefont
  {H.}~\bibnamefont {Gorke}}, \bibinfo {author} {\bibfnamefont
  {J.}~\bibnamefont {Ullrich}}, \bibinfo {author} {\bibfnamefont
  {I.}~\bibnamefont {Schlichting}}, \bibinfo {author} {\bibfnamefont
  {S.}~\bibnamefont {Herrmann}}, \bibinfo {author} {\bibfnamefont
  {G.}~\bibnamefont {Schaller}}, \bibinfo {author} {\bibfnamefont
  {F.}~\bibnamefont {Schopper}}, \bibinfo {author} {\bibfnamefont
  {H.}~\bibnamefont {Soltau}}, \bibinfo {author} {\bibfnamefont {K.-U.}\
  \bibnamefont {K{\"u}hnel}}, \bibinfo {author} {\bibfnamefont
  {R.}~\bibnamefont {Andritschke}}, \bibinfo {author} {\bibfnamefont {C.-D.}\
  \bibnamefont {Schr{\"o}ter}}, \bibinfo {author} {\bibfnamefont
  {F.}~\bibnamefont {Krasniqi}}, \bibinfo {author} {\bibfnamefont
  {M.}~\bibnamefont {Bott}}, \bibinfo {author} {\bibfnamefont {S.}~\bibnamefont
  {Schorb}}, \bibinfo {author} {\bibfnamefont {D.}~\bibnamefont {Rupp}},
  \bibinfo {author} {\bibfnamefont {M.}~\bibnamefont {Adolph}}, \bibinfo
  {author} {\bibfnamefont {T.}~\bibnamefont {Gorkhover}}, \bibinfo {author}
  {\bibfnamefont {H.}~\bibnamefont {Hirsemann}}, \bibinfo {author}
  {\bibfnamefont {G.}~\bibnamefont {Potdevin}}, \bibinfo {author}
  {\bibfnamefont {H.}~\bibnamefont {Graafsma}}, \bibinfo {author}
  {\bibfnamefont {B.}~\bibnamefont {Nilsson}}, \bibinfo {author} {\bibfnamefont
  {H.~N.}\ \bibnamefont {Chapman}}, \ and\ \bibinfo {author} {\bibfnamefont
  {J.}~\bibnamefont {Hajdu}},\ }\href@noop {} {\bibfield  {journal} {\bibinfo
  {journal} {Nature}\ }\textbf {\bibinfo {volume} {470}},\ \bibinfo {pages}
  {78} (\bibinfo {year} {2011})}\BibitemShut {NoStop}%
\bibitem [{\citenamefont {Gerchberg}\ and\ \citenamefont
  {Saxton}(1972)}]{Gerchberg:72}%
  \BibitemOpen
  \bibfield  {author} {\bibinfo {author} {\bibfnamefont {R.~W.}\ \bibnamefont
  {Gerchberg}}\ and\ \bibinfo {author} {\bibfnamefont {W.~O.}\ \bibnamefont
  {Saxton}},\ }\href@noop {} {\bibfield  {journal} {\bibinfo  {journal} {Optik
  (Stuttgart)}\ }\textbf {\bibinfo {volume} {35}} (\bibinfo {year}
  {1972})}\BibitemShut {NoStop}%
\bibitem [{\citenamefont {Fienup}(1982)}]{Fienup:82}%
  \BibitemOpen
  \bibfield  {author} {\bibinfo {author} {\bibfnamefont {J.~R.}\ \bibnamefont
  {Fienup}},\ }\href {\doibase 10.1364/AO.21.002758} {\bibfield  {journal}
  {\bibinfo  {journal} {Appl. Opt.}\ }\textbf {\bibinfo {volume} {21}},\
  \bibinfo {pages} {2758} (\bibinfo {year} {1982})}\BibitemShut {NoStop}%
\bibitem [{\citenamefont {Bauschke}\ \emph {et~al.}(2003)\citenamefont
  {Bauschke}, \citenamefont {Combettes},\ and\ \citenamefont
  {Luke}}]{Bauschke:03}%
  \BibitemOpen
  \bibfield  {author} {\bibinfo {author} {\bibfnamefont {H.~H.}\ \bibnamefont
  {Bauschke}}, \bibinfo {author} {\bibfnamefont {P.~L.}\ \bibnamefont
  {Combettes}}, \ and\ \bibinfo {author} {\bibfnamefont {D.~R.}\ \bibnamefont
  {Luke}},\ }\href {\doibase 10.1364/JOSAA.20.001025} {\bibfield  {journal}
  {\bibinfo  {journal} {J. Opt. Soc. Am. A}\ }\textbf {\bibinfo {volume}
  {20}},\ \bibinfo {pages} {1025} (\bibinfo {year} {2003})}\BibitemShut
  {NoStop}%
\bibitem [{\citenamefont {Luke}(2005)}]{Luke:05}%
  \BibitemOpen
  \bibfield  {author} {\bibinfo {author} {\bibfnamefont {D.~R.}\ \bibnamefont
  {Luke}},\ }\href {http://stacks.iop.org/0266-5611/21/i=1/a=004} {\bibfield
  {journal} {\bibinfo  {journal} {Inverse Problems}\ }\textbf {\bibinfo
  {volume} {21}},\ \bibinfo {pages} {37} (\bibinfo {year} {2005})}\BibitemShut
  {NoStop}%
\bibitem [{\citenamefont {Marchesini}(2007)}]{marchesini:07}%
  \BibitemOpen
  \bibfield  {author} {\bibinfo {author} {\bibfnamefont {S.}~\bibnamefont
  {Marchesini}},\ }\href@noop {} {\bibfield  {journal} {\bibinfo  {journal}
  {Review of scientific instruments}\ }\textbf {\bibinfo {volume} {78}},\
  \bibinfo {pages} {011301} (\bibinfo {year} {2007})}\BibitemShut {NoStop}%
\bibitem [{\citenamefont {Sayre}(1952)}]{sayre:52}%
  \BibitemOpen
  \bibfield  {author} {\bibinfo {author} {\bibfnamefont {D.}~\bibnamefont
  {Sayre}},\ }\href@noop {} {\bibfield  {journal} {\bibinfo  {journal} {Acta
  Crystallographica}\ }\textbf {\bibinfo {volume} {5}},\ \bibinfo {pages} {843}
  (\bibinfo {year} {1952})}\BibitemShut {NoStop}%
\bibitem [{\citenamefont {Ikeda}\ and\ \citenamefont {Kono}(2012)}]{Ikeda:12}%
  \BibitemOpen
  \bibfield  {author} {\bibinfo {author} {\bibfnamefont {S.}~\bibnamefont
  {Ikeda}}\ and\ \bibinfo {author} {\bibfnamefont {H.}~\bibnamefont {Kono}},\
  }\href {\doibase 10.1364/OE.20.003375} {\bibfield  {journal} {\bibinfo
  {journal} {Opt. Express}\ }\textbf {\bibinfo {volume} {20}},\ \bibinfo
  {pages} {3375} (\bibinfo {year} {2012})}\BibitemShut {NoStop}%
\bibitem [{\citenamefont {Thibault}\ and\ \citenamefont
  {Guizar-Sicairos}(2012)}]{thibault:12}%
  \BibitemOpen
  \bibfield  {author} {\bibinfo {author} {\bibfnamefont {P.}~\bibnamefont
  {Thibault}}\ and\ \bibinfo {author} {\bibfnamefont {M.}~\bibnamefont
  {Guizar-Sicairos}},\ }\href@noop {} {\bibfield  {journal} {\bibinfo
  {journal} {New Journal of Physics}\ }\textbf {\bibinfo {volume} {14}},\
  \bibinfo {pages} {063004} (\bibinfo {year} {2012})}\BibitemShut {NoStop}%
\bibitem [{\citenamefont {Paxman}\ \emph {et~al.}(1992)\citenamefont {Paxman},
  \citenamefont {Schulz},\ and\ \citenamefont {Fienup}}]{paxman:92}%
  \BibitemOpen
  \bibfield  {author} {\bibinfo {author} {\bibfnamefont {R.~G.}\ \bibnamefont
  {Paxman}}, \bibinfo {author} {\bibfnamefont {T.~J.}\ \bibnamefont {Schulz}},
  \ and\ \bibinfo {author} {\bibfnamefont {J.~R.}\ \bibnamefont {Fienup}},\
  }\href {\doibase 10.1364/JOSAA.9.001072} {\bibfield  {journal} {\bibinfo
  {journal} {J. Opt. Soc. Am. A}\ }\textbf {\bibinfo {volume} {9}},\ \bibinfo
  {pages} {1072} (\bibinfo {year} {1992})}\BibitemShut {NoStop}%
\bibitem [{\citenamefont {Nishino}\ \emph {et~al.}(2003)\citenamefont
  {Nishino}, \citenamefont {Miao},\ and\ \citenamefont {Ishikawa}}]{Nishino03}%
  \BibitemOpen
  \bibfield  {author} {\bibinfo {author} {\bibfnamefont {Y.}~\bibnamefont
  {Nishino}}, \bibinfo {author} {\bibfnamefont {J.}~\bibnamefont {Miao}}, \
  and\ \bibinfo {author} {\bibfnamefont {T.}~\bibnamefont {Ishikawa}},\ }\href
  {\doibase 10.1103/PhysRevB.68.220101} {\bibfield  {journal} {\bibinfo
  {journal} {Phys. Rev. B}\ }\textbf {\bibinfo {volume} {68}},\ \bibinfo
  {pages} {220101} (\bibinfo {year} {2003})}\BibitemShut {NoStop}%
\bibitem [{\citenamefont {Miao}\ \emph {et~al.}(2005)\citenamefont {Miao},
  \citenamefont {Nishino}, \citenamefont {Kohmura}, \citenamefont {Johnson},
  \citenamefont {Song}, \citenamefont {Risbud},\ and\ \citenamefont
  {Ishikawa}}]{Miao05}%
  \BibitemOpen
  \bibfield  {author} {\bibinfo {author} {\bibfnamefont {J.}~\bibnamefont
  {Miao}}, \bibinfo {author} {\bibfnamefont {Y.}~\bibnamefont {Nishino}},
  \bibinfo {author} {\bibfnamefont {Y.}~\bibnamefont {Kohmura}}, \bibinfo
  {author} {\bibfnamefont {B.}~\bibnamefont {Johnson}}, \bibinfo {author}
  {\bibfnamefont {C.}~\bibnamefont {Song}}, \bibinfo {author} {\bibfnamefont
  {S.~H.}\ \bibnamefont {Risbud}}, \ and\ \bibinfo {author} {\bibfnamefont
  {T.}~\bibnamefont {Ishikawa}},\ }\href {\doibase
  10.1103/PhysRevLett.95.085503} {\bibfield  {journal} {\bibinfo  {journal}
  {Phys. Rev. Lett.}\ }\textbf {\bibinfo {volume} {95}},\ \bibinfo {pages}
  {085503} (\bibinfo {year} {2005})}\BibitemShut {NoStop}%
\bibitem [{\citenamefont {Huang}\ \emph {et~al.}(2010)\citenamefont {Huang},
  \citenamefont {Nelson}, \citenamefont {Steinbrener}, \citenamefont {Kirz},
  \citenamefont {Turner},\ and\ \citenamefont {Jacobsen}}]{Huang:10}%
  \BibitemOpen
  \bibfield  {author} {\bibinfo {author} {\bibfnamefont {X.}~\bibnamefont
  {Huang}}, \bibinfo {author} {\bibfnamefont {J.}~\bibnamefont {Nelson}},
  \bibinfo {author} {\bibfnamefont {J.}~\bibnamefont {Steinbrener}}, \bibinfo
  {author} {\bibfnamefont {J.}~\bibnamefont {Kirz}}, \bibinfo {author}
  {\bibfnamefont {J.~J.}\ \bibnamefont {Turner}}, \ and\ \bibinfo {author}
  {\bibfnamefont {C.}~\bibnamefont {Jacobsen}},\ }\href {\doibase
  10.1364/OE.18.026441} {\bibfield  {journal} {\bibinfo  {journal} {Opt.
  Express}\ }\textbf {\bibinfo {volume} {18}},\ \bibinfo {pages} {26441}
  (\bibinfo {year} {2010})}\BibitemShut {NoStop}%
\bibitem [{\citenamefont {Salha}(2014)}]{Salha:14}%
  \BibitemOpen
  \bibfield  {author} {\bibinfo {author} {\bibfnamefont {S.}~\bibnamefont
  {Salha}},\ }\emph {\bibinfo {title} {Inference from Incomplete Data in
  Coherent Diffraction Imaging}},\ \href@noop {} {Ph.D. thesis},\ \bibinfo
  {school} {UCLA} (\bibinfo {year} {2014}),\ \bibinfo {note}
  {http://www.escholarship.org/uc/item/75x1988b}\BibitemShut {NoStop}%
\bibitem [{\citenamefont {He}\ \emph {et~al.}(2015)\citenamefont {He},
  \citenamefont {Sharma},\ and\ \citenamefont {Cossairt}}]{He:15}%
  \BibitemOpen
  \bibfield  {author} {\bibinfo {author} {\bibfnamefont {K.}~\bibnamefont
  {He}}, \bibinfo {author} {\bibfnamefont {M.~K.}\ \bibnamefont {Sharma}}, \
  and\ \bibinfo {author} {\bibfnamefont {O.}~\bibnamefont {Cossairt}},\ }\href
  {\doibase 10.1364/OE.23.030904} {\bibfield  {journal} {\bibinfo  {journal}
  {Opt. Express}\ }\textbf {\bibinfo {volume} {23}},\ \bibinfo {pages} {30904}
  (\bibinfo {year} {2015})}\BibitemShut {NoStop}%
\bibitem [{\citenamefont {Rudin}\ \emph {et~al.}(1992)\citenamefont {Rudin},
  \citenamefont {Osher},\ and\ \citenamefont {Fatemi}}]{rudin:92}%
  \BibitemOpen
  \bibfield  {author} {\bibinfo {author} {\bibfnamefont {L.~I.}\ \bibnamefont
  {Rudin}}, \bibinfo {author} {\bibfnamefont {S.}~\bibnamefont {Osher}}, \ and\
  \bibinfo {author} {\bibfnamefont {E.}~\bibnamefont {Fatemi}},\ }\href@noop {}
  {\bibfield  {journal} {\bibinfo  {journal} {Physica D: Nonlinear Phenomena}\
  }\textbf {\bibinfo {volume} {60}},\ \bibinfo {pages} {259} (\bibinfo {year}
  {1992})}\BibitemShut {NoStop}%
\bibitem [{\citenamefont {Boyd}\ \emph {et~al.}(2011)\citenamefont {Boyd},
  \citenamefont {Parikh}, \citenamefont {Chu}, \citenamefont {Peleato},\ and\
  \citenamefont {Eckstein}}]{boyd:11}%
  \BibitemOpen
  \bibfield  {author} {\bibinfo {author} {\bibfnamefont {S.}~\bibnamefont
  {Boyd}}, \bibinfo {author} {\bibfnamefont {N.}~\bibnamefont {Parikh}},
  \bibinfo {author} {\bibfnamefont {E.}~\bibnamefont {Chu}}, \bibinfo {author}
  {\bibfnamefont {B.}~\bibnamefont {Peleato}}, \ and\ \bibinfo {author}
  {\bibfnamefont {J.}~\bibnamefont {Eckstein}},\ }\href@noop {} {\bibfield
  {journal} {\bibinfo  {journal} {Foundations and Trends{\textregistered} in
  Machine Learning}\ }\textbf {\bibinfo {volume} {3}},\ \bibinfo {pages} {1}
  (\bibinfo {year} {2011})}\BibitemShut {NoStop}%
\bibitem [{\citenamefont {Netrapalli}\ \emph {et~al.}(2015)\citenamefont
  {Netrapalli}, \citenamefont {Jain},\ and\ \citenamefont
  {Sanghavi}}]{netrapalli:15}%
  \BibitemOpen
  \bibfield  {author} {\bibinfo {author} {\bibfnamefont {P.}~\bibnamefont
  {Netrapalli}}, \bibinfo {author} {\bibfnamefont {P.}~\bibnamefont {Jain}}, \
  and\ \bibinfo {author} {\bibfnamefont {S.}~\bibnamefont {Sanghavi}},\
  }\href@noop {} {\bibfield  {journal} {\bibinfo  {journal} {Signal Processing,
  IEEE Transactions on}\ }\textbf {\bibinfo {volume} {63}},\ \bibinfo {pages}
  {4814} (\bibinfo {year} {2015})}\BibitemShut {NoStop}%
\bibitem [{\citenamefont {Weller}\ \emph {et~al.}(2015)\citenamefont {Weller},
  \citenamefont {Pnueli}, \citenamefont {Divon}, \citenamefont {Radzyner},
  \citenamefont {Eldar},\ and\ \citenamefont {Fessler}}]{weller:15}%
  \BibitemOpen
  \bibfield  {author} {\bibinfo {author} {\bibfnamefont {D.~S.}\ \bibnamefont
  {Weller}}, \bibinfo {author} {\bibfnamefont {A.}~\bibnamefont {Pnueli}},
  \bibinfo {author} {\bibfnamefont {G.}~\bibnamefont {Divon}}, \bibinfo
  {author} {\bibfnamefont {O.}~\bibnamefont {Radzyner}}, \bibinfo {author}
  {\bibfnamefont {Y.~C.}\ \bibnamefont {Eldar}}, \ and\ \bibinfo {author}
  {\bibfnamefont {J.~A.}\ \bibnamefont {Fessler}},\ }\href@noop {} {\bibfield
  {journal} {\bibinfo  {journal} {Computational Imaging, IEEE Transactions on}\
  }\textbf {\bibinfo {volume} {1}},\ \bibinfo {pages} {247} (\bibinfo {year}
  {2015})}\BibitemShut {NoStop}%
\bibitem [{\citenamefont {Fienup}(1993)}]{fienup:93}%
  \BibitemOpen
  \bibfield  {author} {\bibinfo {author} {\bibfnamefont {J.~R.}\ \bibnamefont
  {Fienup}},\ }\href@noop {} {\bibfield  {journal} {\bibinfo  {journal}
  {Applied optics}\ }\textbf {\bibinfo {volume} {32}},\ \bibinfo {pages} {1737}
  (\bibinfo {year} {1993})}\BibitemShut {NoStop}%
\bibitem [{\citenamefont {Soille}()}]{Soille:99}%
  \BibitemOpen
  \bibfield  {author} {\bibinfo {author} {\bibfnamefont {P.}~\bibnamefont
  {Soille}},\ }\href@noop {} {\emph {\bibinfo {title} {Morphological Image
  Analysis: Principles and Applications}}}\BibitemShut {NoStop}%
\bibitem [{\citenamefont {Parikh}\ and\ \citenamefont
  {Boyd}(2014)}]{parikh:14}%
  \BibitemOpen
  \bibfield  {author} {\bibinfo {author} {\bibfnamefont {N.}~\bibnamefont
  {Parikh}}\ and\ \bibinfo {author} {\bibfnamefont {S.~P.}\ \bibnamefont
  {Boyd}},\ }\href@noop {} {\bibfield  {journal} {\bibinfo  {journal}
  {Foundations and Trends in optimization}\ }\textbf {\bibinfo {volume} {1}},\
  \bibinfo {pages} {127} (\bibinfo {year} {2014})}\BibitemShut {NoStop}%
\bibitem [{\citenamefont {Maia}\ \emph {et~al.}(2010)\citenamefont {Maia},
  \citenamefont {Ekeberg}, \citenamefont {Van Der~Spoel},\ and\ \citenamefont
  {Hajdu}}]{maia:10}%
  \BibitemOpen
  \bibfield  {author} {\bibinfo {author} {\bibfnamefont {F.~R.}\ \bibnamefont
  {Maia}}, \bibinfo {author} {\bibfnamefont {T.}~\bibnamefont {Ekeberg}},
  \bibinfo {author} {\bibfnamefont {D.}~\bibnamefont {Van Der~Spoel}}, \ and\
  \bibinfo {author} {\bibfnamefont {J.}~\bibnamefont {Hajdu}},\ }\href@noop {}
  {\bibfield  {journal} {\bibinfo  {journal} {Journal of applied
  crystallography}\ }\textbf {\bibinfo {volume} {43}},\ \bibinfo {pages} {1535}
  (\bibinfo {year} {2010})}\BibitemShut {NoStop}%
\bibitem [{\citenamefont {Berman}\ \emph {et~al.}(2002)\citenamefont {Berman},
  \citenamefont {Battistuz}, \citenamefont {Bhat}, \citenamefont {Bluhm},
  \citenamefont {Bourne}, \citenamefont {Burkhardt}, \citenamefont {Feng},
  \citenamefont {Gilliland}, \citenamefont {Iype}, \citenamefont {Jain},
  \citenamefont {Fagan}, \citenamefont {Marvin}, \citenamefont {Padilla},
  \citenamefont {Ravichandran}, \citenamefont {Schneider}, \citenamefont
  {Thanki}, \citenamefont {Weissig}, \citenamefont {Westbrook},\ and\
  \citenamefont {Zardecki}}]{berman:02}%
  \BibitemOpen
  \bibfield  {author} {\bibinfo {author} {\bibfnamefont {H.~M.}\ \bibnamefont
  {Berman}}, \bibinfo {author} {\bibfnamefont {T.}~\bibnamefont {Battistuz}},
  \bibinfo {author} {\bibfnamefont {T.~N.}\ \bibnamefont {Bhat}}, \bibinfo
  {author} {\bibfnamefont {W.~F.}\ \bibnamefont {Bluhm}}, \bibinfo {author}
  {\bibfnamefont {P.~E.}\ \bibnamefont {Bourne}}, \bibinfo {author}
  {\bibfnamefont {K.}~\bibnamefont {Burkhardt}}, \bibinfo {author}
  {\bibfnamefont {Z.}~\bibnamefont {Feng}}, \bibinfo {author} {\bibfnamefont
  {G.~L.}\ \bibnamefont {Gilliland}}, \bibinfo {author} {\bibfnamefont
  {L.}~\bibnamefont {Iype}}, \bibinfo {author} {\bibfnamefont {S.}~\bibnamefont
  {Jain}}, \bibinfo {author} {\bibfnamefont {P.}~\bibnamefont {Fagan}},
  \bibinfo {author} {\bibfnamefont {J.}~\bibnamefont {Marvin}}, \bibinfo
  {author} {\bibfnamefont {D.}~\bibnamefont {Padilla}}, \bibinfo {author}
  {\bibfnamefont {V.}~\bibnamefont {Ravichandran}}, \bibinfo {author}
  {\bibfnamefont {B.}~\bibnamefont {Schneider}}, \bibinfo {author}
  {\bibfnamefont {N.}~\bibnamefont {Thanki}}, \bibinfo {author} {\bibfnamefont
  {H.}~\bibnamefont {Weissig}}, \bibinfo {author} {\bibfnamefont {J.~D.}\
  \bibnamefont {Westbrook}}, \ and\ \bibinfo {author} {\bibfnamefont
  {C.}~\bibnamefont {Zardecki}},\ }\href@noop {} {\bibfield  {journal}
  {\bibinfo  {journal} {Acta Crystallographica Section D: Biological
  Crystallography}\ }\textbf {\bibinfo {volume} {58}},\ \bibinfo {pages} {899}
  (\bibinfo {year} {2002})}\BibitemShut {NoStop}%
\bibitem [{\citenamefont {Maia}(2012)}]{maia:12}%
  \BibitemOpen
  \bibfield  {author} {\bibinfo {author} {\bibfnamefont {F.~R.}\ \bibnamefont
  {Maia}},\ }\href@noop {} {\bibfield  {journal} {\bibinfo  {journal} {Nature
  methods}\ }\textbf {\bibinfo {volume} {9}},\ \bibinfo {pages} {854} (\bibinfo
  {year} {2012})}\BibitemShut {NoStop}%
\end{thebibliography}


%

\end{document}